\documentclass[prb,twocolumn,10pt,aps,showpacs,floatfix,amsmath,amsfonts,letterpaper]{revtex4-2}

\usepackage{stix} 
\usepackage{hyperref} 
\usepackage{graphicx} 

\newcommand{\ee}{\mathrm{e}}
\newcommand{\ii}{\mathrm{i}}
\newcommand{\dd}{\mathrm{d}}

\newcommand{\rv}{\boldsymbol{r}}
\newcommand{\Rv}{\boldsymbol{R}}

\newcommand{\scH}{\mathcal{H}}
\newcommand{\scZ}{\mathcal{Z}}
\newcommand{\scT}{\mathcal{T}}
\newcommand{\scS}{\mathcal{S}}

\newcommand{\xuv}{\hat{\boldsymbol{x}}}
\newcommand{\yuv}{\hat{\boldsymbol{y}}}
\newcommand{\zuv}{\hat{\boldsymbol{z}}}

\newcommand{\coord}{\mathfrak{z}}

\makeatletter
\def\beq{\@ifstar{\@ifnextchar[{\@beqslabel}{\@beqsnolabel}}%
{\@ifnextchar[{\@beqlabel}{\@beqnolabel}}}
\def\@beqlabel[#1]{\begin{equation}\label{#1}}
\def\@beqnolabel{\begin{equation}}
\def\@beqslabel[#1]{\begin{equation*}\label{#1}}
\def\@beqsnolabel{\begin{equation*}}
\def\eeq{\@ifstar{\end{equation*}}{\end{equation}}}
\makeatother

\newcommand{\refeq}[1]{Eq.~(\ref{#1})}

\newcommand{\refcites}[1]{Refs.~\cite{#1}}
\newcommand{\reffig}[1]{Fig.~\ref{#1}}
\newcommand{\reffigand}[2]{Figs.~\ref{#1} and \ref{#2}}
\newcommand{\refsec}[1]{Section~\ref{#1}}

\newcommand{\punc}[1]{\,{\text{#1}}}
\newcommand{\sub}[1]{_{\text{#1}}}

\newcommand{\super}[1]{^{\text{#1}}}

\usepackage{color}

\begin{document}

\title{Quantum Kasteleyn transition}

\author{Stephen Powell}
\affiliation{School of Physics and Astronomy, The University of Nottingham, Nottingham, NG7 2RD, United Kingdom}

\date{\today}

\begin{abstract}
Dimer models arise as effective descriptions in a variety of physical contexts, and provide paradigmatic examples of systems subject to strong local constraints. Here we present a quantum version of the venerable Kasteleyn model, which has an unusual phase transition from a dimer solid to a U(1) liquid. We show how the phase structure of the quantum model can be understood in terms of the quantum mechanics of one-dimensional strings and determine the exact value of the critical coupling. By constructing effective models to describe the properties of these strings, we calculate properties such as the dimer--dimer correlation function in the neighborhood of the transition. We also discuss the full phase structure of the model, in the ground state and at nonzero temperature.
\end{abstract}

\maketitle

\section{Introduction}
\label{sec:introduction}

Dimer models, in which the elementary degrees of freedom are hard-core objects on the links of a lattice, are examples of strongly constrained systems showing interesting collective phenomena. Both classical \cite{Henley2010} and quantum \cite{MoessnerRaman2011} dimer models provide simple examples of systems exhibiting phenomena such as topological order, fractionalization, and unconventional phase transitions.

The statistical mechanics of classical dimers was studied by Kasteleyn \cite{Kasteleyn1961} and by Temperley and Fisher \cite{Temperley1961}, who found exact results for two-dimensional (2D) lattices. Interest in quantum dimer models was initiated by Anderson's suggestion \cite{Anderson1987}, in the context of high-temperature superconductivity, that frustrated quantum antiferromagnets could exhibit phases in which nearby pairs of spins form resonating singlets. Rokhsar and Kivelson \cite{Rokhsar1988} subsequently introduced the quantum dimer model as an effective description, treating the dimers as the elementary degrees of freedom and writing down the simplest possible quantum Hamiltonian in terms of them.

Dimer models can exhibit a wide array of different ordered phases, sometimes referred to as ``valence bond solids'' \cite{MoessnerRaman2011}, which can be characterized using conventional order parameters. They are also known to exhibit more unconventional ``dimer liquid'' phases, which arise due to the strong correlations inherent in close-packed dimer configurations (see \refsec{SecClassicalDimerModel}) and are closely related to classical and quantum spin liquids in spin models \cite{Savary2017}. They include classical dimer models on bipartite lattices, which host Coulomb phases \cite{Henley2010}, described by effective gauge theories and with monomers, defects in the close-packing constraint, acting as deconfined charges.

Transitions between these two types of phases differ in various ways from conventional phase transitions between ordered and disordered phases. They include cases where symmetry is spontaneously broken at the transition \cite{Alet2005,Alet2006}, which do not fit into the standard Landau classification, because of the strong correlations on the liquid side. They also include transitions where no symmetry is spontaneously broken \cite{Wilkins2019,Wilkins2020,Desai2021}, and the transition is instead characterized through the loss of topological order, or, more concretely, by confinement of monomers.

The (classical) Kasteleyn transition \cite{Kasteleyn1963} is an example of the latter class, with the unusual property that the model is exactly solvable. The transition occurs in the classical dimer model on the honeycomb lattice with a potential-energy term favoring dimers occupying a particular subset of the links. In this work, we introduce a quantum analogue of the Kasteleyn model and show that it has a similar transition, but at zero temperature, driven by quantum fluctuations. Although we cannot solve the model exactly, we can precisely determine the location of the transition, because of the particular nature of the ``ordered'' phase. This therefore provides a rare of example of a quantum phase transition where the critical coupling can be calculated exactly.

Previous extensions to Kasteleyn's work include a study by Bhattacharjee et al.\ \cite{Bhattacharjee1983} of a related model on a three-dimensional (3D) analogue of the brick lattice, formed by deleting alternate vertical bonds of the cubic lattice. A 3D variant of the Kasteleyn problem has been studied in spin ice \cite{Jaubert2008,SpinIceCQ}, where the constraints on the dimer configurations are replaced by the ``ice rule''. Closely related quantum models have also been studied, in the context of dimers \cite{HerzogArbeitman2019} and quantum spin ice \cite{Wan2012}.

In the remainder of this section, we define the classical and quantum models. In \refsec{SecClassicalKasteleyn}, we then review the relevant aspects of the classical Kasteleyn problem, including the argument that determines its exact critical temperature. Most of our main results are presented in \refsec{SecQuantumKasteleyn}, where we discuss the quantum Kasteleyn transition. We briefly discuss the full phase structure of the model in \refsec{SecDenseStringPhase} before concluding in \refsec{SecConclusions}.

\subsection{Classical dimer model}
\label{SecClassicalDimerModel}

In the models we consider, the elementary degrees of freedom are dimers on the links of a lattice, with the number of dimers on each link \(\ell\) restricted to \(d_\ell = 0,1\). A close-packed dimer configuration is one where every site of the lattice has exactly one dimer, i.e., where the number of dimers on links \(\ell\) connected to site \(i\),
\beq[EqDefinen]
n_i = \sum_{\ell \in i} d_\ell\punc,
\eeq
is fixed to \(n_i = 1\) for each \(i\).

\begin{figure}
\includegraphics{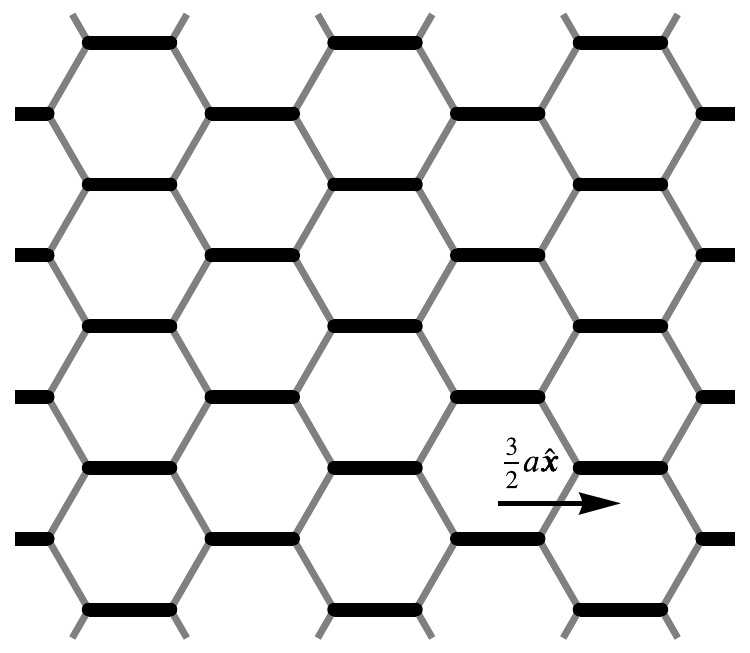}
\caption{Honeycomb lattice with horizontal links highlighted (thick black lines). In the Kasteleyn problem, each occupied horizontal link contributes energy \(-u\). The \(x\) axis is horizontal and the three nearest-neighbor vectors are \(\pm a(\xuv\cos\theta + \yuv\sin\theta)\) for the two sublattices (\(\pm\)), where \(a\) is the nearest-neighbor distance and \(\theta = 0,\frac{2\pi}{3},\frac{4\pi}{3}\). The arrow shows the vector \(\frac{3}{2}a \xuv\), the horizontal projection of the displacement between two adjacent columns of horizontal links.}
\label{FigLattice}
\end{figure}
\begin{figure}
\includegraphics{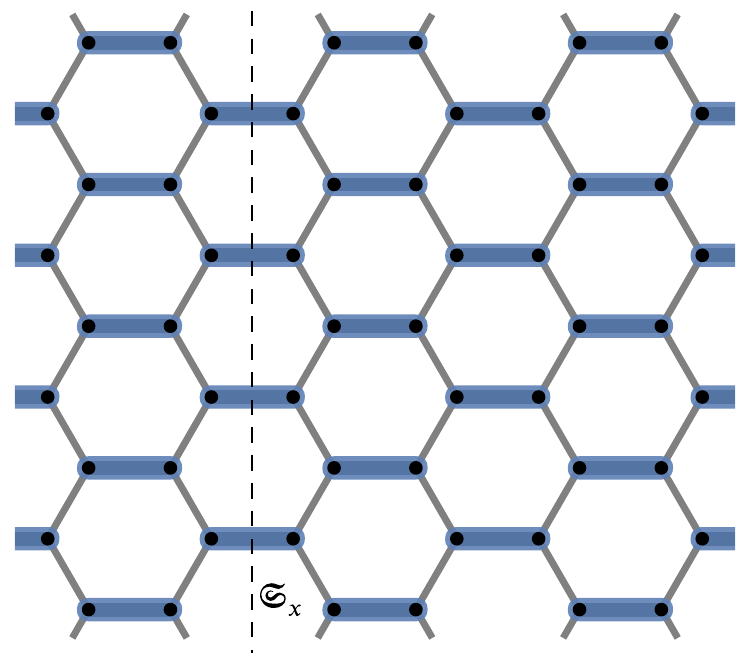}
\caption{Classical ground state of the potential illustrated in \reffig{FigLattice}, where all horizontal links (and only those links) are occupied. There are no possible local rearrangements compatible with the close-packing constraint on the dimers. The dashed vertical line \(\mathfrak{S}_x\) is used to define the horizontal component of the flux, \(\Phi_x\).}
\label{FigStaggered}
\end{figure}

For a classical dimer model, the partition function is given by
\beq[EqPartitionFunction]
\scZ = \sum_{c} \ee^{-V/T}\punc{,}
\eeq
where the sum is over all close-packed configurations and \(V\) is the energy assigned to configuration \(c\). (We set \(k\sub{B}=1\) throughout.) The classical Kasteleyn problem \cite{Kasteleyn1963} concerns dimers on the honeycomb lattice with
\beq[EqKasteleynH]
V = -u N\sub{h}\punc{,}
\eeq
where \(N\sub{h}\) is the number of dimers on horizontal links; see \reffig{FigLattice}. One can instead think of \(N\sub{h}\) as the overlap (i.e., number of coinciding dimers) between each configuration and a fixed reference configuration, shown in \reffig{FigStaggered}, with all horizontal links occupied. This is a valid close-packed configuration, and is hence clearly the unique ground state of \(V\), with the maximal value \(N\sub{h} = N\), equal to the total number of dimers.

\begin{figure}
\includegraphics{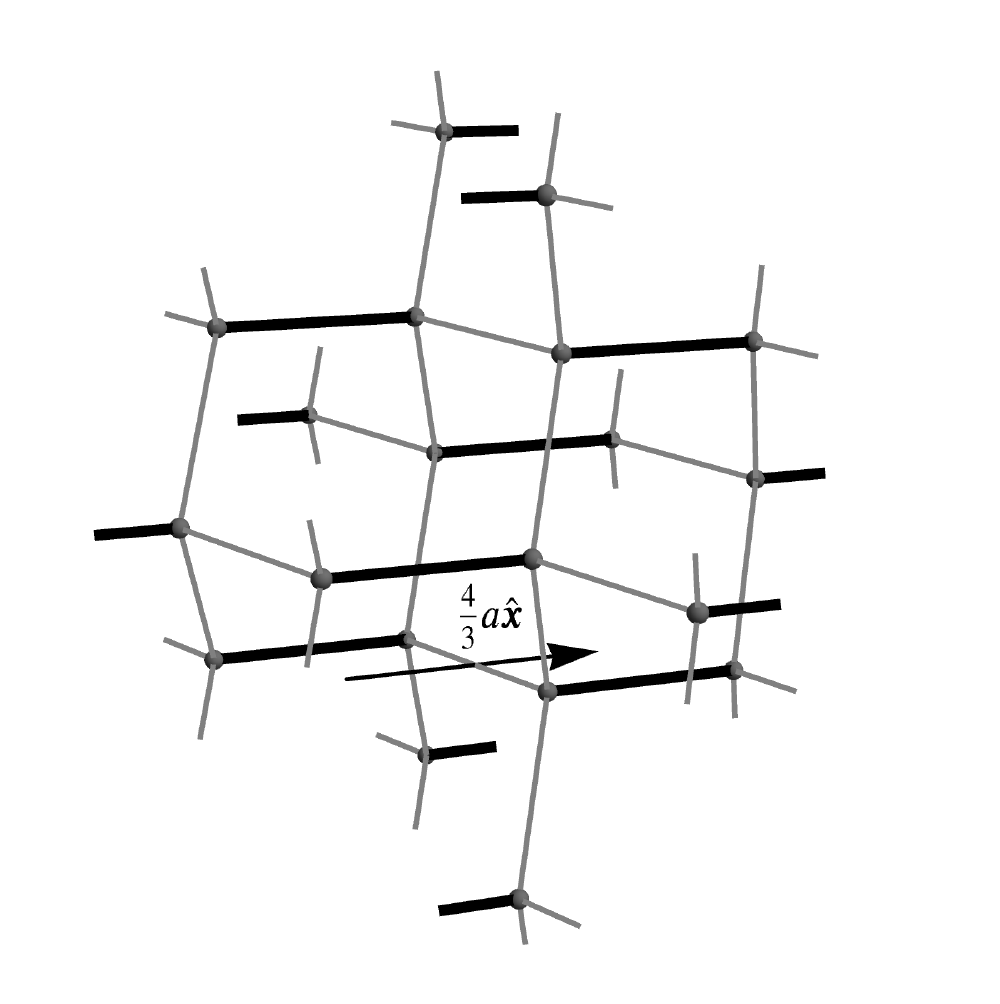}
\caption{Diamond lattice, with horizontal links, i.e., those parallel to the \(x\) axis, highlighted (thick black lines). The four nearest-neighbor vectors are \(\mp a\xuv\) and \(\pm a(\frac{1}{3}\xuv + \frac{\sqrt{8}}{3}\yuv\cos\theta + \frac{\sqrt{8}}{3}\zuv\cos\theta)\) for the two sublattices (\(\pm\)), where \(a\) is the nearest-neighbor distance and \(\theta = 0,\frac{2\pi}{3},\frac{4\pi}{3}\). The arrow shows the vector \(\frac{4}{3}a \xuv\), the horizontal projection of the displacement between two adjacent planes of horizontal links (compare \reffig{FigLattice}).}
\label{FigLatticeDiamond}
\end{figure}

The unusual properties of the Kasteleyn transition result from the fact that, starting from this reference configuration, there are no possible local rearrangements that maintain close packing. Instead, as we describe in \refsec{SecClassicalKasteleyn}, the minimal excitations involve shifting dimers along paths spanning the system, and hence cost unbounded energy in the thermodynamic limit. The problem can be generalized to other lattices, including in higher dimensions, by similarly choosing reference configurations on each, with this same property. For concreteness, we focus here on the honeycomb lattice along with the 3D diamond lattice, illustrated in \reffig{FigLatticeDiamond}, with the potential again applied on horizontal links.

The detailed properties of the model are clearly sensitive to the boundary conditions, which must be chosen to be compatible both with the close-packing constraint and with the reference configuration. To simplify the analysis, we choose one of the periodic lattice vectors along the horizontal direction \(\xuv\) and the others in the plane perpendicular to this. For honeycomb (resp.\ diamond), this can be achieved by choosing \(\Rv\sub{h} = \frac{3}{2}aL \xuv\) (\(\Rv\sub{h} = \frac{4}{3}aL  \xuv\)) as one of the periodic lattice vectors, where \(a\) is the nearest-neighbor distance and \(L\) is divisible by \(2\) (\(3\)). Consider a codimension-\(1\) surface \(\mathfrak{S}_x\) perpendicular to \(\xuv\) that spans the system and passes through a layer of horizontal links, such as the dashed line in \reffig{FigStaggered} or a \(yz\) plane in the diamond lattice. With these definitions, the number of horizontal links through which \(\mathfrak{S}_x\) passes is given by \(W_\perp = N/L\).

We note in passing that \(N\sub{h}\) is related to the horizontal component \(\Phi_x\) of the {\it flux} \cite{Chalker2017} (or ``winding number'' \cite{MoessnerRaman2011}). On both the honeycomb and diamond lattices, this can be defined as
\beq[EqFlux]
\Phi_x = \sum_{\ell \in \mathfrak{S}_x} \left(d_\ell - \frac{1}{\coord}\right)
\eeq
where the sum is over links passing through the surface \(\mathfrak{S}_x\) and \(\coord\) is the coordination number (\(3\) for honeycomb, \(4\) for diamond). For close-packed dimer configurations, the value of \(\Phi_x\) is the same for any choice of the horizontal position of the surface \cite{Chalker2017}; averaging \refeq{EqFlux} over all \(L\) such positions gives \(\Phi_x = \frac{1}{L}\left(N\sub{h} - \frac{N}{\coord}\right)\). The reference configuration, which maximizes the number of dimers on horizontal links, therefore also has maximal \(\Phi_x = \Phi\sub{max} \equiv W_\perp\left(1-\frac{1}{\coord}\right)\). A state that preserves rotational symmetry, such as when \(u = 0\) (assuming no spontaneous symmetry breaking), has \(N\sub{h} = \frac{N}{\coord}\) and hence \(\Phi_x = 0\).

\subsection{Quantum Kasteleyn model}

To construct a quantum dimer model, one defines orthogonal basis vectors \(\lvert c \rangle\) corresponding to each dimer configuration \(c\), and a Hilbert space spanned by the full set. We also define (hard-core bosonic) operators \(b_\ell\) and \(b^\dagger_\ell\) that, respectively, annihilate and create a dimer on link \(\ell\), as well as the dimer number operator \(d_\ell = b_\ell^\dagger b_\ell\). One can restrict to close-packed configurations by projecting into the subspace with eigenvalue \(1\) for the operator \(n_i\), defined by \refeq{EqDefinen}, for each site \(i\). The operators \(b_{\ell}^{(\dagger)}\) do not commute with this projection, but combinations can be constructed that relate different close-packed configurations and hence do commute with it.

\begin{figure}
\includegraphics{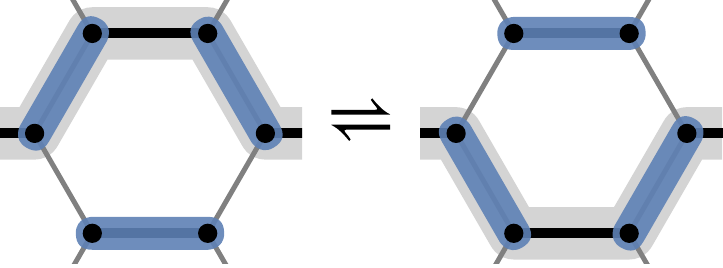}
\vspace{2em}\\
\includegraphics{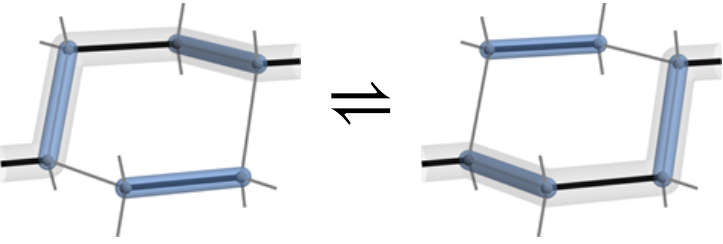}
\caption{Action of the plaquette-flip operator \(\scT_p\) for the honeycomb (top) and diamond (bottom) lattices. In both cases, the effect is to shift three dimers around a hexagonal plaquette (which is nonplanar in the case of diamond). Note that the number of occupied horizontal bonds (thick black lines) is equal to \(1\) before and after the flip, and hence that \(\scT_p\) commutes with \(N\sub{h}\). The light-gray background in both panels illustrates the path of a string excitation (see \refsec{SecClassicalKasteleyn}), representing the difference from a reference configuration with all dimers on horizontal links. Flipping the plaquette causes the string to move, by exchanging the direction taken by the string on two adjacent steps.}
\label{FigFlip}
\end{figure}

Any such kinetic term involves shifting dimers along closed paths of even length. The simplest is of the type introduced by Rokhsar and Kivelson \cite{Rokhsar1988}, which flips dimers around a plaquette \(p\) of the lattice. We denote this \(\scT_p\) and write the Hamiltonian as
\beq[EqHamiltonian]
\scH = -u N\sub{h} - t \sum_{p} \scT_p
\punc{,}
\eeq
where the sum is over plaquettes of a certain type, typically chosen as the smallest plaquettes of even length. For both the honeycomb and diamond lattices, these are hexagons (nonplanar in the case of diamond), and so the operator \(\scT_p\) can be written as
\beq[EqKineticOperator]
\scT_p = b_{\ell_1}^{\phantom{\dagger}} b_{\ell_2}^\dagger b_{\ell_3}^{\phantom{\dagger}} b_{\ell_4}^\dagger b_{\ell_5}^{\phantom{\dagger}} b_{\ell_6}^\dagger + \text{h.c.}
\punc{,}
\eeq
where \(\ell_{1\cdots6}\) are the six links comprising the plaquette. These operations are illustrated in \reffig{FigFlip}.
Our main focus will be the ground-state phase structure of this model, as a function of the dimensionless parameter \(t/u\).

\begin{figure}
\includegraphics{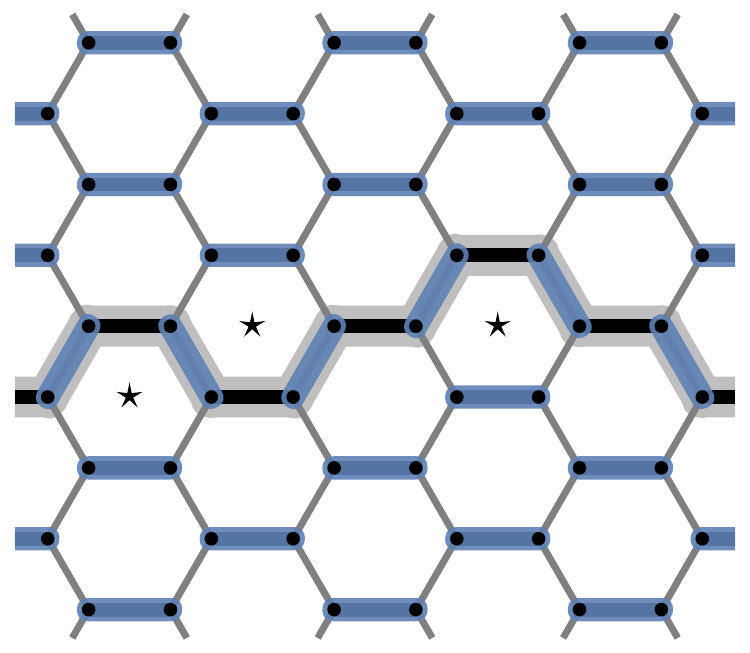}
\caption{Example of a string on the honeycomb lattice, where a set of dimers (indicated with a light-gray background) have been shifted compared to the reference configuration shown in \reffig{FigStaggered}. Such strings, which span the system in the horizontal direction, are the minimal rearrangements compatible with the close-packing constraint. Note that plaquettes are flippable (marked with a star, \(\star\)) when two adjacent steps of the string are in opposite directions.}
\label{FigString}
\end{figure}

By inspection of \reffig{FigFlip}, it is clear that flipping dimers around a plaquette conserves the number of occupied horizontal links, and hence that \(\scT_p\) commutes with \(N\sub{h}\). The same is therefore true when moving dimers around any closed loop that can be constructed by combining plaquettes, but not for topologically nontrivial loops, i.e., those that span the boundaries. Shifting dimers around such loops can change \(N\sub{h}\), as illustrated in \reffig{FigString}.
(Since \(N\sub{h}\) is related to \(\Phi_x\), the conservation of the number of horizontal dimers under local rearrangements is in fact a particular instance of the general conservation of flux in dimer models \cite{MoessnerRaman2011}.)

\section{Classical Kasteleyn problem}
\label{SecClassicalKasteleyn}

We first review the phase structure of the classical Kasteleyn problem. While Kasteleyn \cite{Kasteleyn1963} solved the classical model exactly, by expressing the partition function \refeq{EqPartitionFunction} in terms of a determinant, it is more instructive for our purposes to argue from the high- and low-temperature limits.

We start with the high-temperature limit \(T/u = \infty\), where all close-packed dimer configurations have equal weight. Even in this case, the thermal state is nontrivial, because of the correlations inherent in this set of configurations. The resulting state, which survives to finite \(T/u\), is referred to as a {\it Coulomb phase} \cite{Henley2010} and is a ``dimer liquid'', in the sense that there is no spontaneous symmetry breaking but the positions of the dimers are strongly correlated.

In the opposite limit, \(T/u = 0\), the system enters its ground state, equal to the reference configuration with all dimers horizontal (and hence maximal flux \(\Phi_x\)). To determine the effect of small nonzero temperature, we consider the excitations of the system. Since, by design, there are no local rearrangements of the reference configuration that preserve the close-packing constraint, any excitation that does so must span the boundaries. It is easily verified that the minimal excitation in fact involves shifting dimers along a path that spans the system once in the horizontal direction, moving the dimers from horizontal links to links in other directions, as illustrated in \reffigand{FigString}{FigStringDiamond}. We refer to such an object as a {\it string}.

\begin{figure*}
\includegraphics[width=0.48\textwidth]{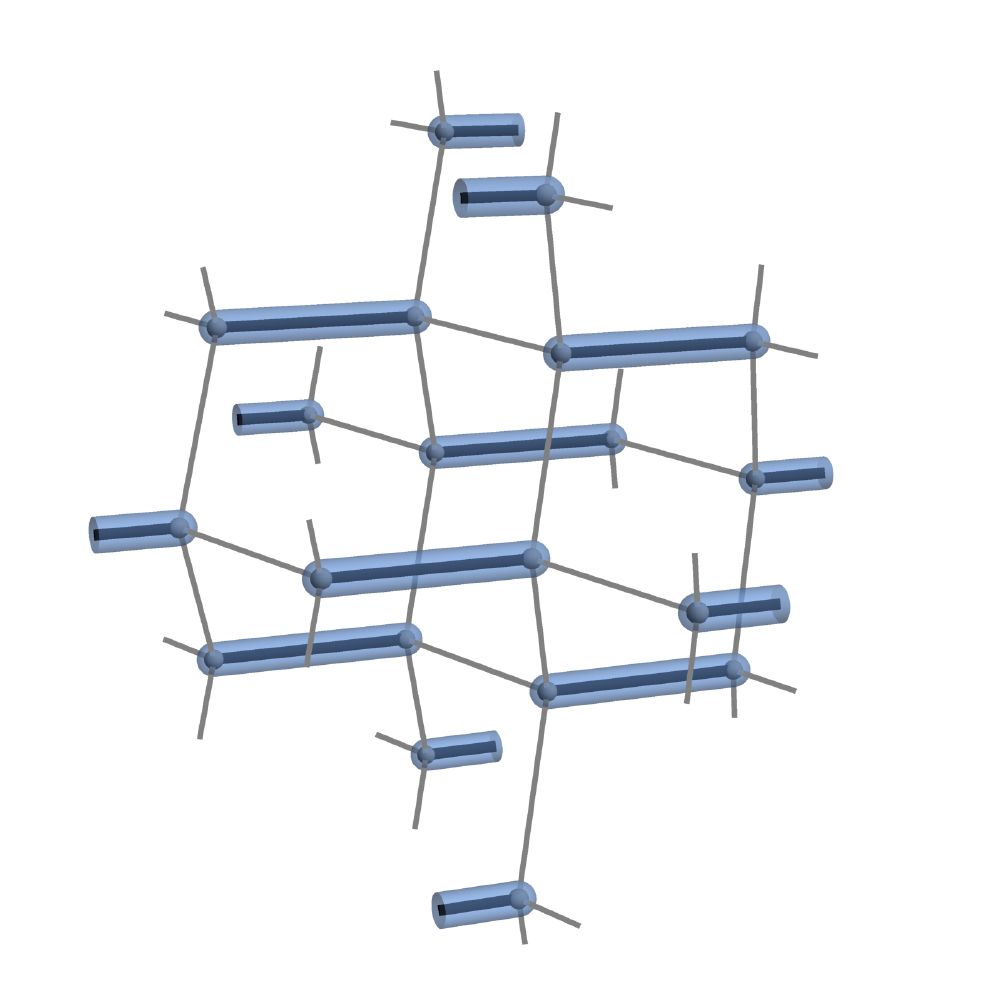}
\includegraphics[width=0.48\textwidth]{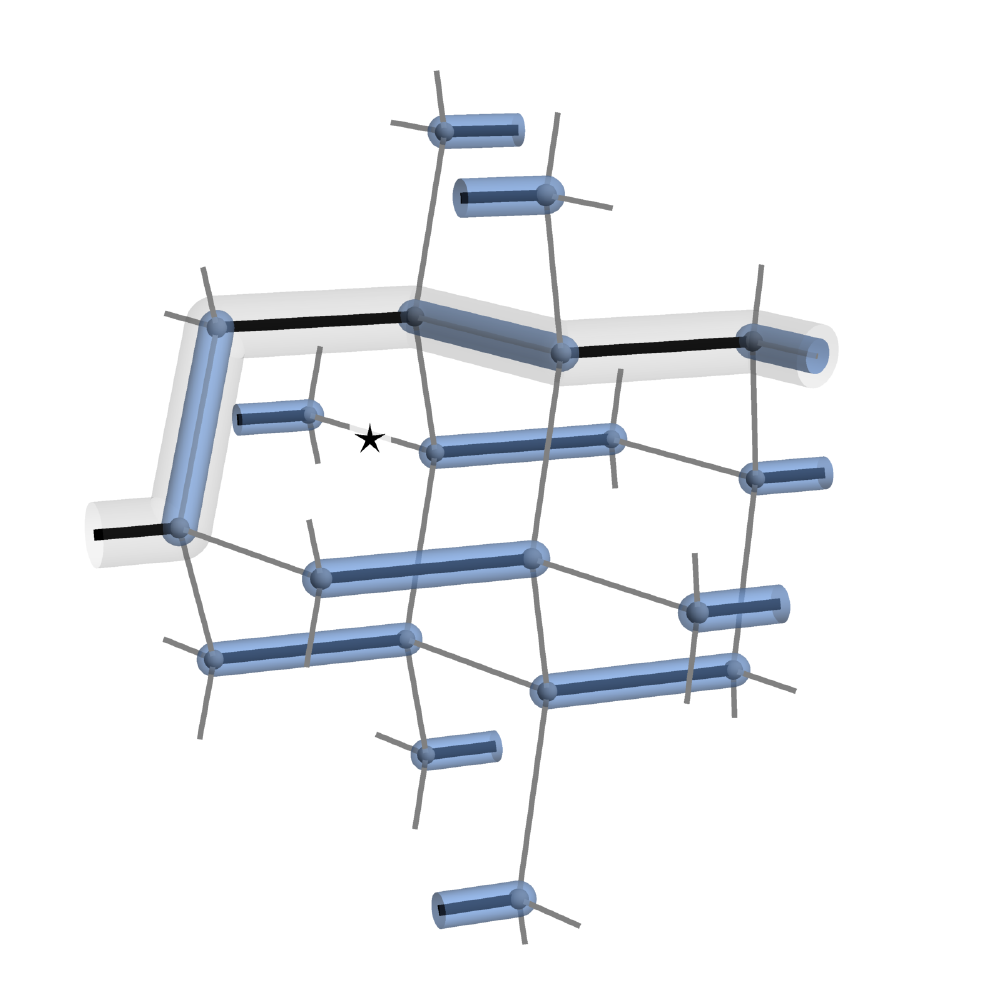}
\caption{Left: Classical ground state of the potential shown in \reffig{FigLatticeDiamond}, with all horizontal links occupied. As in \reffig{FigStaggered}, there are no local rearrangements of dimers consistent with the close-packing constraint. Right: Example of a string on the diamond lattice, where a set of dimers have been shifted that spans the system in the horizontal direction, as on the honeycomb lattice, \reffig{FigString}. Two consecutive steps that are in different directions again produce a flippable plaquette, marked with a star (\(\star\)).}
\label{FigStringDiamond}
\end{figure*}

Each step in the path of a string, from one horizontal link to the next, involves a horizontal displacement \(\frac{3}{2}a\) for honeycomb (see \reffig{FigLattice}) and \(\frac{4}{3}a\) for diamond (see \reffig{FigLatticeDiamond}). A string that spans the system once in the horizontal direction therefore reduces \(N\sub{h}\) by \(L\). We denote the number of strings by \(n\sub{s}\), and so we have \(N\sub{h} = N - n\sub{s}L\).

From \refeq{EqKasteleynH}, introducing a string increases the configuration energy \(V\) by \(+uL\). It may nonetheless reduce the \emph{free} energy, and hence be thermodynamically favorable, since it also makes a positive contribution to the entropy. This entropy arises from the multiple possible choices for the direction taken at each step on the path, of which there are \(q = \coord - 1\), i.e., \(q=2\) (up or down) on the honeycomb lattice and \(q=3\) on the diamond lattice. In both cases, the entropy is \(\ln q\) per unit length, giving a change in the free energy of \(\Delta F = (u - T \ln q)L\). An inserted string reduces the free energy, and is hence thermodynamically favorable, when \(T > T\sub{K}\), where \(T\sub{K} = u / \ln q\) gives the exact transition temperature for the classical Kasteleyn model on the honeycomb and diamond lattices. (The same logic can be applied starting from similar reference configurations on other lattices \cite{Bhattacharjee1983}, although in most cases the different possible choices at each step are not equivalent, and so the calculation is more involved \cite{Oakes2016,HerzogArbeitman2019}.)

For finite \(L\), only strings returning to the same position, i.e., with zero net displacement in the transverse directions, are allowed. This restriction reduces the entropy only by an amount of order \(\ln L\), and hence does not affect \(T\sub{K}\). On the honeycomb lattice, for example, the vertical displacement follows a binomial distribution, and so the proportion of strings with zero net displacement is \(\propto L^{-1/2}\), giving a reduction of the entropy of order \(\ln L\).

As the temperature increases above \(T\sub{K}\), string excitations become thermodynamically favorable, and \(n\sub{s}\) increases from zero. For \(n\sub{s} \ll W_\perp\), the total entropy is given by a sum of single-string contributions, \(S(n\sub{s})\approx n\sub{s}L \ln q\), while for larger \(n\sub{s}\) steric interactions (i.e., the fact that strings cannot overlap) reduce the entropy per string.

The energy with \(n\sub{s}\) strings is exactly \(V = - u (N - n\sub{s} L) = -u L (W_\perp - n\sub{s})\), and so the free energy in an ensemble with fixed string number \(n\sub{s}\) is
\beq
F(n\sub{s}) = -uL W_\perp + u L n\sub{s} - T S(n\sub{s})\punc.
\eeq
In the thermodynamic limit, \(L,W_\perp\rightarrow\infty\), the string density \(\rho\sub{s} = n\sub{s}/W_\perp\) takes the value that minimizes \(F(n\sub{s})\), which is zero for \(T < T\sub{K}\) and nonzero for \(T > T\sub{K}\). The value of \(\rho\sub{s}\) at the minimum for \(T > T\sub{K}\) is determined by the (positive) higher-order terms in \(F(n\sub{s})\) resulting from the entropy reduction due to interactions.

\begin{figure}
\includegraphics{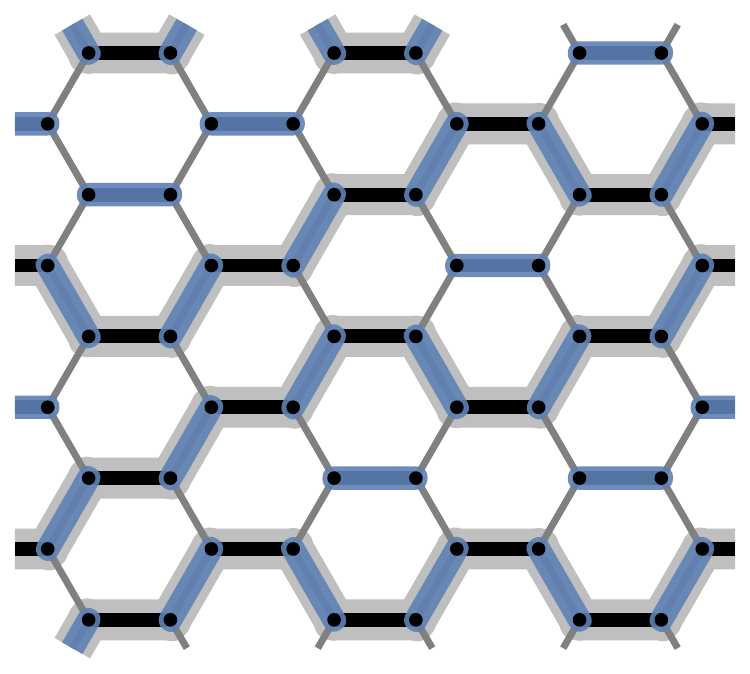}
\caption{Example configuration of dimers on the honeycomb lattice, mapped to a set of strings superimposed on the reference configuration. Given a fixed reference configuration, in this case the one with all dimers horizontal, there is a one-to-one mapping from close-packed dimer configurations to configurations of zero or more nonoverlapping strings.}
\label{FigRandom}
\end{figure}

With a nonzero density of strings, the system is a fluctuating dimer liquid, continuously connected to the point at \(T/u = +\infty\), and forming a Coulomb phase in the entire region \(T > T\sub{K}\). There is a one-to-one mapping from any configuration of dimers to a collection of (nonoverlapping) strings superimposed on the reference configuration, as illustrated in \reffig{FigRandom}, and so, far from the transition, one can view the system as containing a dense set of strings, with \(\rho\sub{s} = 1 - N\sub{h}/N\) of order unity.
The mean density of strings increases continuously with \(T/u\) in the Coulomb phase until the proportion of horizontal dimers reaches \(N\sub{h}/N = 1/\coord\) (and hence the flux \(\Phi_x \propto N\sub{h} - {N}/{\coord}\) reaches zero), by symmetry, at \(T/u = +\infty\).

For large but finite \(L\), one can distinguish the two phases at \(T \gtrless T\sub{K}\) by how the distribution of \(N\sub{h}\) (or equivalently of \(n\sub{s}\)) scales with \(L\). Strings are exponentially suppressed with \(L\) for \(T < T\sub{K}\), where they cost positive free energy scaling linearly with \(L\). In the Coulomb phase for \(T > T\sub{K}\), the variance of \(N\sub{h}\) scales as \(L^d\) \cite{Henley2010}, as for a typical extensive quantity in the thermodynamic limit. In the thermodynamic limit, there is a phase transition between these two asymptotic behaviors, but there is no spontaneous symmetry breaking on either side of the transition.

A critical theory for the Kasteleyn transition can be found by mapping the \(d\)-dimensional classical system to a \((d-1)\)-dimensional quantum system, with the strings becoming the world lines for bosons \cite{Jaubert2008,SpinIceCQ}. Note that the resulting theory is not the standard proliferation transition in the inverted-XY universality class \cite{Banks1977,Dasgupta1981}, because the strings are directed and no local closed loops are possible.

\section{Quantum Kasteleyn problem}
\label{SecQuantumKasteleyn}

We now turn to the quantum Kasteleyn model at zero temperature. The Hamiltonian \(\scH\) defined in \refeq{EqHamiltonian}, and indeed any local Hamiltonian, cannot change the number of strings \(n\sub{s}\), because creating or annihilating a string involves shifting dimers along a path that spans the system. (The string number is a quantity conserved by a topological constraint, rather than by a symmetry.) The Hilbert space therefore splits into sectors of fixed \(n\sub{s}\), and the ground-state energy \(E\sub{gs}(n\sub{s})\) in each will play the same role as \(F(n\sub{s})\) in the classical case.

The limit \(t/u = 0\) coincides with the low-temperature limit of the classical model; the ground state is given by the reference configuration with all dimers on horizontal links. This state has no flippable hexagons, and so the plaquette-flip operator has no effect on this configuration, which remains an exact eigenstate, with energy \(E\sub{gs}(0) = -u L W_\perp\), for any \(t\).

As in the classical case, introducing a single string into the system costs potential energy \(+uL\). For nonzero \(t\), it also causes plaquettes along the length of the string to become flippable, as illustrated in \reffig{FigString}. Flipping these causes the string to move, deforming along its length; see \reffig{FigFlip}. The string therefore delocalizes, giving a negative contribution to the kinetic energy. Since the Hamiltonian is local, the kinetic energy of an isolated string will generally also be proportional to \(L\) in the thermodynamic limit.

It is therefore possible for the net energy change of introducing a string into the system to be negative, as long as the kinetic energy overcomes the potential energy associated with the string. As in the classical case, the criterion for proliferation of strings, i.e., whether the global ground state of \(\scH\) occurs for \(n\sub{s} = 0\) or \(n\sub{s} > 0\), can be determined by considering the asymptotic behavior at small \(n\sub{s}\). We therefore begin by considering the properties of a single string.

\subsection{Effective single-string Hamiltonian}

Because \(\scH\) conserves the number of strings, one can write an effective Hamiltonian \(\scH\super{string}\) in the single-string Hilbert space. For both the honeycomb and diamond lattices, the degrees of freedom describing a string are the \(q\) possible directions that it takes at each step, with \(q=2\) for honeycomb and \(q=3\) for diamond. We define \(\lvert \alpha \rangle_i\), for \(\alpha = 1, 2, \dotsc, q\), as a basis state representing a step in direction \(\alpha\) at position \(i\) along the string (with periodic boundary conditions).

The kinetic term in \(\scH\) produces a term that allows the direction on adjacent steps to be swapped provided that they are different (see \reffigand{FigFlip}{FigString}). The single-string Hamiltonian is therefore
\beq
\scH\super{string} = \sum_{i=1}^L \left(u - t \scT\super{string}_{i,i+1} \right)\punc,
\eeq
where
\beq
\scT\super{string}_{i,j} = \sum^q_{\substack{\alpha,\beta=1\\\alpha\neq\beta}} \lvert \alpha \rangle_i\lvert \beta \rangle_{j} \langle \beta \rvert_{i} \langle \alpha \rvert_{j}
\eeq
is an operator that permutes the flavors at sites \(i\) and \(j\), or gives zero if they are the same.

This model has permutation symmetry under exchange of any two directions, corresponding to real-space reflections, as well as \(\mathrm{U}(1)^q\) symmetry under diagonal unitary transformations, but not full \(\mathrm{SU}(q)\) symmetry (because \(\alpha = \beta\) is not included in the sum).

To apply periodic boundary conditions along the direction of the string, one should also enforce the constraint that the total number of steps in each direction is equal, so that the net transverse displacement is zero. This constraint commutes with \(\scH\sub{string}\), which conserves the total number of each type of step, and so, provided the ground state does not break permutation symmetry, this constraint has no effect.

\subsubsection{Honeycomb lattice}
\label{SecQKHoneycomb}

For the honeycomb lattice, \(q=2\), and the single-string Hamiltonian can be rewritten in terms of Pauli operators \(\sigma_i^\pm = \frac{1}{2}\left(\sigma_i^x \pm \sigma_i^y\right)\), as
\beq
\scH\super{string}_{q=2} = \sum_i \left[ u - t \left(\sigma^+_i \sigma^-_{i+1} + \sigma^-_i \sigma^+_{i+1}\right)\right]\punc,
\eeq
where \(\sigma^+_i = (\sigma^-_i)^\dagger = \lvert 2 \rangle_i\langle 1 \rvert_i\). This is the Hamiltonian for a spin-\(\frac{1}{2}\) XX chain, which can trivially be solved using a Jordan--Wigner transformation. (An identical effective model was in fact found to describe strings in quantum spin ice on the checkerboard lattice \cite{Wan2012}.)

Defining fermionic operators \(c_i\) by
\beq[EqJW]
\begin{aligned}
\sigma_i^+ &= c_i\prod_{j < i}(1 - 2c_j^\dagger c_j)\\
\sigma_i^z &= 2c_i^\dagger c_i - 1
\punc,
\end{aligned}
\eeq
the Hamiltonian becomes
\beq
\scH\super{string}_{q=2} = \sum_i \left[ u - t \left( c_{i+1}^\dagger c_i + c_i^\dagger c_{i+1} \right) \right]\punc.
\eeq
Transforming to momentum space gives free fermions with dispersion \(\epsilon_k = -2t \cos k\), where \(-\pi < k \le \pi\) and \(kL/\pi\) should be even (odd) when the fermion number is odd (even) \cite{DePasquale2008}.

In the ground state, all \(L/2\) single-particle states with \(\epsilon_k < 0\) are occupied, giving total energy
\beq[EqHoneycombGSEnergy]
\begin{aligned}
E\super{string}_{q=2} &= u L - 2t \sum_{n = -n\sub{F}}^{n\sub{F}} \cos \frac{2\pi n}{L}\\
&= u L - \frac{2t}{\sin \frac{\pi}{L}}\punc,
\end{aligned}
\eeq
where \(n\sub{F} = \frac{L}{4} - \frac{1}{2}\).
The net vertical displacement of the string is \(\sum_i \sigma^z_i = \sum_i (2 c_i^\dagger c_i - 1) = 0\) in this state, and so the periodic boundary conditions are satisfied.

Taking the thermodynamic limit gives ground-state energy per site of
\beq[EqEgsq=2]
\frac{E\super{string}_{q=2}}{L} = u - \frac{2}{\pi}t
\punc.
\eeq
The energetic contribution of a single string is therefore positive for \(u > \frac{2}{\pi} t\), which implies that the vacuum of strings is the ground state for \(t/u < (t/u)\sub{c} = \frac{\pi}{2}\) on the honeycomb lattice.

\subsubsection{Diamond lattice}

For diamond, where \(q=3\), it is no longer possible to express the string degrees of freedom in terms of spin-\(\frac{1}{2}\) operators, and hence the Jordan--Wigner transformation cannot be used to diagonalize the Hamiltonian.

Instead, one can estimate the ground-state energy using mean-field theory, by taking as the trial state an equal-weight superposition of all permutations \(p\),
\beq
\lvert \text{mf}\rangle \propto \sum_p \lvert p \rangle\punc.
\eeq
For any \(p\), \(\scT\super{string}_{i,i+1}\lvert p \rangle\) gives another permutation appearing in the sum, unless the two directions at \(i\) and \(i+1\) are the same, in which case it gives zero. The expectation value \(\langle \text{mf}\rvert \scT\super{string}_{i,i+1} \lvert \text{mf} \rangle\) is therefore given by the fraction of permutations where the two are different, which is \((q-1)/q\) in the thermodynamic limit. The expectation value is therefore
\beq
\frac{\langle \text{mf}\rvert \scH \lvert \text{mf} \rangle}{L} = u - \frac{q-1}{q} t
\punc,
\eeq
which gives an upper bound on the true ground-state energy per site. For \(q=2\), \(E\super{string}_{q=2}/L \le u - \frac{1}{2}t\), which is consistent with and gives a reasonable variational estimate to \refeq{EqEgsq=2}. For \(q=3\), the upper bound is
\beq
\frac{E\super{string}_{q=3}}{L} \le u - \frac{2}{3}t\punc.
\eeq

\begin{figure}
\includegraphics{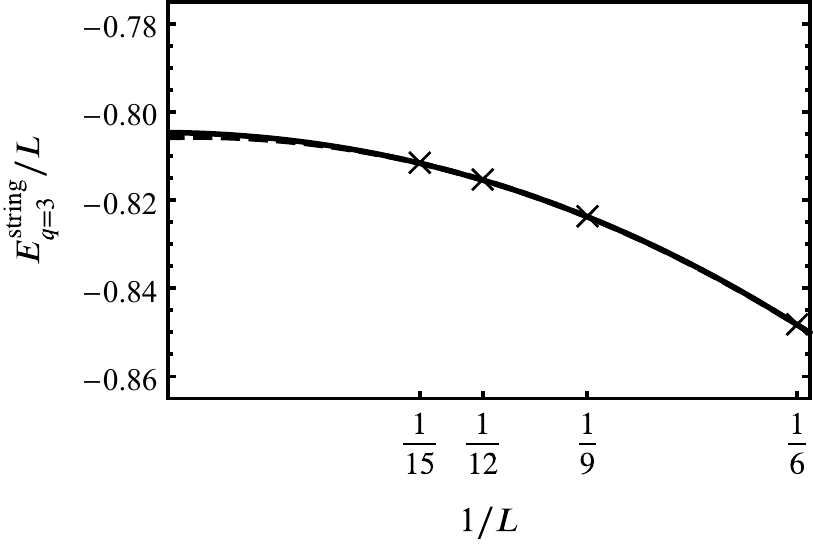}
\caption{Exact-diagonalization results for the ground-state energy per site of a single string on the diamond lattice, for \(t=1\) and \(u=0\), plotted versus inverse system size \(1/L\). The solid and dashed lines show cubic and quadratic fits extrapolated to \(L=\infty\), giving an estimate of \(E\super{string}_{q=3}/L = -0.805(5)\).}
\label{FigEDResults}
\end{figure}

Results from exact diagonalization and an extrapolation to large \(L\) are shown in \reffig{FigEDResults}. Expressing the ground-state energy per site as
\beq
\frac{E\super{string}_{q=3}}{L} = u - \gamma t\punc,
\eeq
these results give an estimate of \(\gamma = 0.805(5)\).
The vacuum of strings is the ground state on the diamond lattice for \(t/u < (t/u)\sub{c} = \gamma^{-1}\).

\subsection{Quantum Kasteleyn transition}
\label{SecQuantumKasteleynTransition}

The ground-state energy \(E\super{string}\) of the single-string Hamiltonian gives the difference between the ground-state energies in the one-string and zero-string sectors, \(E\sub{gs}(1) - E\sub{gs}(0)\). For \(t/u > (t/u)\sub{c}\), where this difference is negative, the overall ground state of \(\scH\) must be in a sector with nonzero string number \(n\sub{s}\). Since the ground state for \(t/u = 0\) certainly has \(n\sub{s}=0\), this implies a phase transition at some value of \(t/u\), between the string vacuum and a phase with nonzero string density.

If we assume that the effective interactions between strings are repulsive, at least in the low-density limit, then the terms in \(E\sub{gs}(n\sub{s})\) that are of higher order in \(n\sub{s}\) are positive. (Such an assumption is at least highly plausible, because of the hard-core repulsion between strings, as in the classical case.) The global minimum of \(E\sub{gs}(n\sub{s})\) then increases continuously from zero as \(t/u\) increases above \((t/u)\sub{c}\), giving a continuous quantum phase transition at critical value \((t/u)\sub{c}\).

\subsubsection{Statistics of string width}
\label{SecQKStringWidth}

For \(t/u\) just above \((t/u)\sub{c}\), we expect the physics to be described, at least over certain length scales, by a set of independent strings, each in the ground state of \(\scH\super{string}\). To determine the properties of the system, we first use the microscopic model for the honeycomb lattice, where it is solvable, before addressing the general case using an effective long-wavelength description.

In terms of the Pauli operators defined for the honeycomb lattice in \refsec{SecQKHoneycomb}, the net vertical displacement between horizontal positions \(i\) and \(j\) (with \(j > i\)) is
\beq
Y_{i,j} = \sum_{r=i}^{j-1}\sigma^z_r\punc,
\eeq
in units of \(\frac{\sqrt{3}}{2}a\). Applying the Jordan--Wigner transformation, \refeq{EqJW}, this is given by \(Y_{i,j} = j - i - 2 N_{i,j}\),
where
\beq
N_{i,j} = \sum_{r=i}^{j-1}c_r^\dagger c_r
\eeq
is the total occupation number on sites \(r\) with \(i \le r < j\). In the ground state of \(\scH\super{string}_{q=2}\), all single-fermion states with \(\epsilon_k < 0\) are occupied, giving a mean density of \(\frac{1}{2}\), and so \(\langle Y_{i,j} \rangle = 0\), as expected by symmetry.

The variance of \(Y_{i,j}\) can be calculated using Wick's theorem, since \(\scH\super{string}_{q=2}\) is quadratic. A standard calculation gives number--number correlations exhibiting Friedel oscillations (with \(2k\sub{F}=\pi\)),
\begin{multline}
\langle c^\dagger_{r+x} c_{r+x} c^\dagger_{r} c_{r} \rangle - \langle c^\dagger_{r+x} c_{r+x}\rangle\langle c^\dagger_{r} c_{r} \rangle\\
= \begin{cases}
\frac{1}{4}&\text{for \(x = 0\)}\\
0 & \text{for even \(x \neq 0\)}\\
-\pi^{-2}x^{-2}&\text{for odd \(x\),}
\end{cases}
\end{multline}
for \(x \ll L\). Summing over \(x\) then gives the result
\beq[EqYij2]
\langle Y_{i,j}^2 \rangle \approx \frac{2}{\pi^2} \ln (j-i)
\eeq
for \(1 \ll j-i \ll L\). This contrasts with the classical case, where a string behaves as a random walk and the variance is instead proportional to \(j-i\). (Note that we have ignored boundary conditions in the transverse directions in this calculations. This is clearly justified, in both the quantum and classical cases, for an isotropic system with large \(L\).)

Instead of using the microscopic model, it is possible to write down an effective continuum field theory of a single string. In 2D, we consider a coarse-grained picture in which the vertical displacement of the string is given by a real function \(\varphi\) of the horizontal position \(x\), taken as continuous. This picture can be generalized to \(d\) spatial dimensions by describing the displacement transverse to the horizontal direction by a \((d-1)\)-component vector \(\varphi_\mu\).

Let \(\pi_\mu(x)\) be the conjugate momentum operator, with canonical commutation relations
\beq[EqCommutator]
\left[\varphi_\mu(x),\pi_\nu(x')\right] = \ii \delta_{\mu\nu} \delta(x-x')\punc,
\eeq
where \(\mu\) and \(\nu\) run over the \(d-1\) directions transverse to the string. The effective Hamiltonian can then be expressed as
\beq[EqStringEff]
\scH\super{string,eff} = \int \dd x\,\left\{\frac{1}{2}\kappa \lvert \boldsymbol{\pi}(x)\rvert^2 + \frac{1}{2}\lambda \lvert \boldsymbol{\varphi}'(x)\rvert^2\right\}\punc,
\eeq
plus higher-order terms, where \(\kappa\) and \(\lambda\) are real parameters.

No mass term (i.e., \(\lvert \boldsymbol{\varphi}(x)\rvert^2\)) is allowed in the string Hamiltonian because of translation symmetry in the transverse directions. Similarly, the higher-order terms must be expressed only in terms of the derivative \(\boldsymbol{\varphi}'\) and the momentum \(\boldsymbol{\pi}\). This implies, by power counting, that they are RG-irrelevant at the Gaussian fixed point represented by \refeq{EqStringEff}, and hence that the field theory is quadratic in the long-wavelength limit. The irrelevance of higher-order terms has the effect that there is no coupling between the different components of \(\varphi\) for \(d>2\).

In principle, a periodic function of \(\varphi_\mu(x)\) could also be added to \(\scH\super{string,eff}\), since the transverse position of the string takes discrete values in the microscopic model. If such a term were relevant, which would require sufficiently small \(\kappa/\lambda\), this would describe a phase where the string was ``flat''. Stabilising such a phase would require additional terms in the original dimer Hamiltonian and would preclude the possibility of a Kasteleyn transition driven by string kinetic energy.

We therefore work with the free field theory of \refeq{EqStringEff}, which can be solved by expressing \(\boldsymbol{\varphi}\) and \(\boldsymbol{\pi}\) in terms of creation and annihilation operators for the modes of the string. This gives for the variance of the displacement
\beq[EqStringWidth]
\left\langle \left[ \varphi_\mu(x) - \varphi_\mu(x') \right]^2 \right\rangle \approx \sqrt{\frac{\kappa}{\lambda}}\frac{1}{\pi}\ln\lvert x-x'\rvert\punc,
\eeq
which is consistent with the microscopic result for \(d=2\), \refeq{EqYij2}. Both expressions are valid only when much smaller than square of the typical separation between strings, given by \(({W_\perp}/{n\sub{s}})^{2/(d-1)}\), and hence constitute intermediate asymptotics \cite{Jaubert2008}.

\subsubsection{Dimer correlations in the single-string limit}

At low string density, the properties of the original dimer model can be expressed in terms of observables for a single string. As an example, we consider the dimer--dimer equal-time correlation function.

For simplicity, consider two horizontal links \(\ell\) and \(\ell'\)  on the honeycomb lattice, separated by displacement \(3a x \xuv + \frac{\sqrt{3}}{2}ay \yuv\), where \(x\) and \(y\) are integers with the same parity. Assuming the string passes through \(\ell\), it will pass through \(\ell'\) if and only if \(Y_{x,0} = y\). A horizontal link is occupied unless a string passes through it, and so the dimer--dimer correlation function obeys
\beq
G(x,y) \equiv \langle (1-d_\ell)(1-d_{\ell'}) \rangle = \frac{1}{W_\perp} \left\langle \delta_{y - Y_{x,0}} \right\rangle\punc,
\eeq
where \(\delta\) is the Kronecker delta. On the right-hand side, \(1/W_\perp\) is the probability that the string passes through \(\ell\), while the expectation value is the conditional probability that it also passes through \(\ell'\). (Correlation functions of \(d_\ell\) for links of other orientations can be expressed similarly.)

Taking the Fourier transform with respect to \(y\) gives
\beq
\tilde{G}(x,k_y) \equiv \sum_y \ee^{-\ii k_y y} G(x,y) = \frac{1}{W_\perp} \left\langle \ee^{-\ii k_y Y_{0,x}} \right\rangle\punc.
\eeq
Again using Wick's theorem, the expectation value can be evaluated to give
\beq
\tilde{G}(x,k_y) = \frac{1}{W_\perp} \ee^{-\frac{1}{2}k_y^2 \langle Y_{0,x}^2\rangle} \approx \frac{1}{W_\perp} x^{-{k_y^2}/{\pi^2}}\punc,
\eeq
using \refeq{EqYij2}, so that (again ignoring transverse boundary conditions) the real-space correlation function is given by
\beq
G(x,y) \approx \frac{1}{W_\perp}\sqrt{\frac{\pi}{4\ln x}} \exp \left(-\frac{\pi^2}{4\ln x}y^2\right)\punc,
\eeq
a Gaussian in \(y\), of width \(\propto \sqrt{\ln x}\).

For nonzero but small string number \(n\sub{s} \ll W_\perp\), one expects that multiple strings will add incoherently to this correlation function, replacing the factor \(1/W_\perp\) by the string density \(\rho\sub{s} = n\sub{s}/W_\perp\).

\section{Dense-string phase}
\label{SecDenseStringPhase}

For \(t/u > (t/u)\sub{c}\), the ground state of \(\scH\) is in a sector with nonzero string density \(\rho\sub{s}\). In this case, the properties of the system (on length scales larger than the typical string separation) are dependent on interactions between strings, and so cannot be inferred directly from the single-string picture of the previous section. To determine the full phase structure in this region would require numerical simulations and is beyond the scope of this work. Instead, we sketch the likely possibilities.

Phases at nonzero string density can be divided into solids (or ``valence bond crystals''), where the dimers form a crystalline arrangement, spontaneously breaking spatial symmetries, and liquids, where they do not. Quantum dimer liquids, as well as their classical analogues \cite{Henley2010}, can also be characterized by phenomena such as deconfinement of monomers and topological order \cite{MoessnerRaman2011}.

In the classical case, the Kasteleyn model has a liquid (Coulomb) phase for all \(T > T\sub{K}\) \cite{Kasteleyn1963} and interactions between dimers are required to produce a solid. In quantum dimer models, by contrast, quantum fluctuations can, and often do, select particular ordered structures, leading to dimer solids. To connect this discussion to more conventional language for dimer models, note (see \refsec{SecClassicalDimerModel}) that the string density \(\rho\sub{s}\) is related to the horizontal flux by \(\Phi_x = \Phi\sub{max} - \rho\sub{s}W_\perp\), and hence that increasing the string density from zero corresponds to reducing the flux from its maximum. In the limit \(t/u=\infty\) the model is rotationally symmetric and so \(\Phi_x = 0\). For \((t/u)\sub{c}< t/u < \infty\), we are therefore interested in states with intermediate flux, \(0 < \Phi_x < \Phi\sub{max}\).

We first consider the 3D diamond lattice. In this case, we can derive some insight from quantum Monte Carlo (QMC) studies of the standard Rokhsar--Kivelson (RK) QDM \cite{Rokhsar1988} on the same lattice by Sikora et al.\ \cite{Sikora2009,Sikora2011}. Instead of the \(u\) term in the Kasteleyn model, this model has a potential term, with coefficient \(\mu\), which counts the number of flippable plaquettes. For \(\mu\) less than a critical value \(\mu\sub{c}\), the ground state was shown to be a dimer solid, referred to as the ``R'' state, while for \(\mu\sub{c} < \mu < t\), it was found to be a quantum dimer liquid. (Here, \(t\) is the coefficient of the kinetic term, and so the solvable ``RK point'' is at \(\mu=t\).)

Significantly, the QMC results give \(\mu\sub{c}/t = 0.75 \pm 0.02\) \cite{Sikora2011}, indicating that at \(\mu=0\), where the RK QDM coincides with the point \(u=0\) of our model, the system is in a dimer solid phase. This implies that the quantum Kasteleyn model on the diamond lattice has a dimer solid phase at \(t/u=\infty\), and perhaps also for large finite \(t/u\). The R state has fixed zero flux, and so reducing \(t/u\) might be expected to favor other states, possibly leading to a transition into an intermediate phase (or phases), before the Kasteleyn transition into the string vacuum at \(t/u = (t/u)\sub{c}\). These might include a quantum dimer liquid with continuously varying flux (and string density).

Such a phase would be a deconfined \(\mathrm{U}(1)\) quantum dimer liquid, as in the RK QDM at \(\mu\sub{c} < \mu < t\), described by a gauge field \(\boldsymbol{A}\) and exhibiting emergent electrodynamics \cite{Sikora2009,Sikora2011,Hermele2004}. While no symmetries are spontaneously broken, the potential term in our original Hamiltonian reduces the spatial symmetry of the model (for \(u \neq 0\)), leading to an effective action \(\scS\), in continuous space \(\rv\) and imaginary time \(\tau\),
\beq[EqGaugeAction3D]
\scS = \frac{1}{2}\int \dd^2 \rv \int \dd \tau \, \left( \lvert \partial_\tau \boldsymbol{A} \rvert^2 + c_x^2 B_x^2 +  c_\perp^2 \lvert\boldsymbol{B}_\perp\rvert^2 \right)\punc,
\eeq
where \(\boldsymbol{B} = \boldsymbol{\nabla} \times \boldsymbol{A}\), with different ``speeds of light'' \(c_x \neq c_\perp\) parallel and perpendicular to the \(x\) axis. The ``flux density'' \(\boldsymbol{B}\) is effectively the string density (with a direction assigned arbitrarily along \(\pm\xuv\)) but with the mean string density subtracted to remove a term linear in \(B_x\).

The possibilities are more limited in 2D because, according to an argument by Polyakov \cite{Polyakov1977}, \(\mathrm{U}(1)\) quantum liquids cannot exist as extended phases. Phases of the 2D RK QDM with intermediate flux have previously been considered by \refcites{Fradkin2004,Vishwanath2004}, in the case where a nonzero flux occurs spontaneously as a result of symmetric interactions between dimers. (Their analyses focus on the region near the RK point, \(\mu = t\), whereas our model is effectively far from this point, with \(\mu = 0\).) Here, we briefly sketch the results as they apply to the Kasteleyn model.

\begin{figure}
\includegraphics{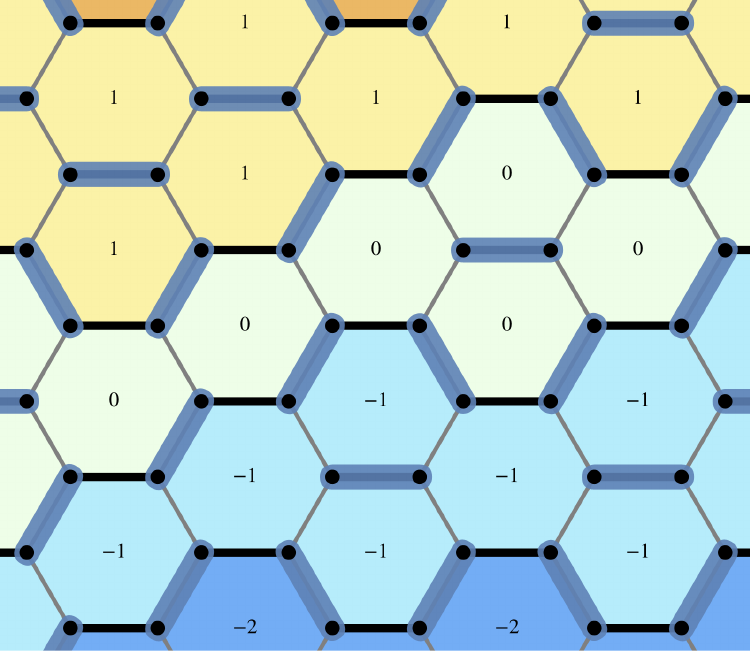}
\caption{Example of the height \(h\) for a configuration of the dimer model on the honeycomb lattice. An integer value \(h\) is assigned to each plaquette such that \(h\) increases by \(1\) when crossing, in the positive \(y\) direction, a horizontal link that is unoccupied or any other link that is occupied. Strings (see \reffig{FigRandom}) therefore act as contours for the height.}
\label{FigHeights}
\end{figure}
A similar coarse-grained action to \refeq{EqGaugeAction3D} can be written in 2D, but with the vector field \(\boldsymbol{A}\) replaced by a scalar ``height'' \(h\) \cite{MoessnerRaman2011}. For our purposes, the simplest way to define the height is to treat each string as a unit step, so that \(h\) is defined on plaquettes and monotonically nondecreasing with \(y\), as illustrated in \reffig{FigHeights}. (This construction is equivalent to the standard mapping from dimers to heights \cite{MoessnerRaman2011}, but with the heights in the fully staggered configuration, \reffig{FigStaggered}, subtracted.) The overall ``tilt'' of the height, i.e., the discontinuity when going around the periodic boundaries in the vertical direction, is given by the string number \(n\sub{s}\).

The local gauge redundancy of \(\boldsymbol{A}\) is replaced in 2D by a global redundancy under uniform shifts of \(h\). Besides the derivative terms analogous to those in \refeq{EqGaugeAction3D}, locking terms of the form \(\cos(2\pi n h)\), with integer \(n\), are therefore also allowed in the action. These terms always lock the height to certain (tilted) configurations, each corresponding to a particular dimer solid. A quantum dimer liquid therefore cannot exist except possibly at isolated points in the phase diagram, where the tilt is incommensurate with the lattice \cite{Fradkin2004,Vishwanath2004}.

\section{Conclusions}
\label{SecConclusions}

This work has introduced a quantum analogue of the Kasteleyn transition in the classical dimer model. As in the classical case, on the ``ordered'' side of the transition, the system is fluctuationless (in the thermodynamic limit) as a consequence of the strong constraints on fluctuations within the set of close-packed dimer configurations. Quantum fluctuations, which play an analogous role to thermal fluctuations in the classical model, can drive a transition to a state with a nonzero density of string excitations. The critical value \((t/u)\sub{c}\) of the ratio of kinetic and potential terms can be calculated exactly on the honeycomb lattice, providing a rare example of a quantum phase transition whose critical coupling can be determined exactly \cite{Sachdev2011}. On the diamond lattice, \((t/u)\sub{c}\) can be expressed in terms of the ground-state energy of a particular 1D quantum model.

The analysis here has focused on the ground state of the quantum Kasteleyn model. At nonzero temperature \(T\), one expects quantum and thermal fluctuations to act in the same direction, and so \((t/u)\sub{c}\) should decrease with \(T\), reaching zero at the transition temperature \(T\sub{K}\) of the classical Kasteleyn model. A (classical) dimer liquid is in principle possible at any \(T>0\) in the quantum model in 2D and 3D, though the quantum dimer solids that likely exist at \(T=0\) would survive up to \(T\) of order of their energy gaps.

While the honeycomb and diamond lattices have been used here for simplicity, the model can be extended to other bipartite lattices by appropriate choice of the potential term, or, equivalently, of the reference configuration. On the square lattice, for example, an appropriate reference configuration has dimers in a staggered arrangement, which maximizes the horizontal component of the flux and hosts analogous string excitations \cite{Oakes2016,HerzogArbeitman2019}.

\acknowledgments

This work was supported by EPSRC Grant No.\ EP/T021691/1.

\end{document}